\documentclass[twocolumn,aps,floatfix,showpacs,prl]{revtex4-2}

\usepackage{amsmath}
\usepackage{amssymb}
\usepackage{esint}
\usepackage{graphicx}
\usepackage{bbold}
\usepackage[normalem]{ulem}
\usepackage{tabularx}
\usepackage{booktabs}
\usepackage{svg}

\usepackage{afterpage}
\usepackage{caption}

\usepackage{setspace}
\usepackage{breakurl}
\usepackage{seqsplit}
\usepackage[hidelinks,colorlinks]{hyperref}
\hypersetup{citecolor=[rgb]{0.25,0.14,0.63}}
\hypersetup{urlcolor=[rgb]{0.25,0.14,0.63}}
\hypersetup{linkcolor=black}

\makeatletter

\usepackage[utf8]{inputenc}
\usepackage{mathrsfs}
\usepackage{xspace}
\usepackage[T1]{fontenc}
\usepackage{dcolumn}
\usepackage{bm}
\usepackage{float}
\usepackage[capitalise]{cleveref}
\usepackage[normalem]{ulem}
\usepackage{color}

\usepackage{esint}

\usepackage{setspace}

\usepackage{breakurl}
\usepackage{seqsplit}
\usepackage[hidelinks,colorlinks]{hyperref}
\hypersetup{citecolor=[rgb]{0.25,0.14,0.63}}
\hypersetup{urlcolor=[rgb]{0.25,0.14,0.63}}
\hypersetup{linkcolor=black}

\begin{document}

\title{Multipartite entanglement swapping with homodyne and single photon detection}
\author{Tarek Assaf}
\author{Josephine Dias}
\email{j.dias@uq.edu.au}
\affiliation{School of Mathematics and Physics, University of Queensland, Brisbane, QLD 4072,  Australia}

\begin{abstract}
Establishing multipartite entanglement across lossy optical networks remains a central challenge for distributed quantum information processing. Here, we consider a non-Gaussian continuous-variable entanglement-swapping protocol combining a symmetric discrete Fourier transform (DFT) interferometer with photon-number-resolving detection. In the ideal limit, single-photon heralding generates a $W_N$-class state across remote modes, with Fourier symmetry rendering all single-click outcomes equivalent up to local phase corrections. We find that conditional pairwise entanglement can exceed that of conventional Gaussian homodyne swapping in weak-to-moderate squeezing and lossy regimes, remaining nonzero beyond the channel-loss thresholds at which Gaussian pairwise entanglement vanishes. These results demonstrate that combining non-Gaussian measurements with symmetric interferometric mixing can improve both the resilience and operational efficiency of multipartite entanglement distribution in lossy quantum networks.
\end{abstract}

\maketitle
\section{Introduction}

Quantum entanglement is a fundamental resource for quantum information processing, underpinning applications in quantum communication, computation, and sensing \cite{horodecki2009quantum}. Establishing entanglement between spatially separated systems is essential for scalable quantum networks and distributed quantum technologies \cite{gisin2007quantum, kimble2008quantum, wehner2018quantum, munro2022designing, azuma2023quantum}.

Distributing entanglement across remote nodes presents significant challenges. Optical transmission loss and environmental noise degrade quantum correlations, limiting the distances over which entanglement can be directly distributed \cite{Pirandola2017PLOB}. As quantum networks scale to accommodate multiple users, approaches that establish entanglement across interconnected links become increasingly important.

Entanglement swapping provides such a mechanism, enabling remote parties that have not directly interacted to share entanglement through joint measurements at an intermediate relay \cite{Zukowski1993}. This process is a central building block of quantum repeaters, enabling long-distance entanglement distribution by connecting shorter entangled links \cite{briegel1998quantum, sangouard2011quantum, munro2015inside, dias2017quantum}. In continuous-variable (CV) systems  \cite{Weedbrook2012, BraunsteinVanLoock2005, Serafini2017, usenko2026continuous}, Gaussian states, linear optics, and homodyne detection provide an accessible framework for implementing this process \cite{loock1999unconditional, hoelscherobermaier2011optimal, polkinghorne1999continuous}. Ottaviani \textit{et al.} extended this approach to multipartite networks, employing an interferometric relay and homodyne measurements to generate symmetric Gaussian entangled states among remote users \cite{Ottaviani2019}.

However, Gaussian entanglement swapping remains sensitive to optical loss and noise, with the resulting entanglement degrading as channel loss and network size increase \cite{namiki2014gaussian}. Here, we focus on star network schemes in which each of $N$ users holds a bipartite entangled state and sends one mode to a central relay, where a joint measurement entangles the modes retained by the users \cite{bose1998multiparticle, Ottaviani2019, roga2023efficient, ishihara2026multipartite, alwehaibi2026improving}. We compare non-Gaussian single-photon detection (SPD) to homodyne measurement for multipartite entanglement swapping.  We consider a symmetric discrete Fourier transform (DFT) relay that enables acceptance of all symmetry-equivalent single-photon heralding outcomes. Our results demonstrate improved conditional entanglement in selected lossy regimes relative to Gaussian swapping, while the DFT architecture enhances heralding probability and entanglement yield without reducing conditional entanglement.


\section{Background}
\subsection{Multipartite continuous-variable entanglement swapping}
\label{sec:background_ottaviani}

Entanglement swapping enables two or more remote systems to become entangled through a joint measurement on modes that have interacted only at an intermediate relay \cite{Zukowski1993}. In continuous-variable (CV) systems, Gaussian states and measurements provide a natural implementation using squeezed optical modes, linear interferometers, and homodyne detection. Ottaviani \textit{et al.} extended this principle to the multipartite setting, allowing an arbitrary number of users to share
a permutation-symmetric Gaussian state following a multipartite CV Bell measurement at a central relay \cite{Ottaviani2019}.

Consider $N$ users, each initially possessing an identical bipartite state $\rho_{A_iB_i}$. Each user retains mode $B_i$ while transmitting mode $A_i$ to the central relay. For two-mode squeezed-vacuum (TMSV) resources, the input state may be written as

\begin{equation}
    |\psi_{\mathrm{TMSV}}\rangle_{A_iB_i}
    =
    \sqrt{1-\lambda^2}
    \sum_{n=0}^{\infty}
    \lambda^n
    |n\rangle_{A_i}|n\rangle_{B_i},
    \qquad
    \lambda=\tanh r ,
    \label{eq:tmsv}
\end{equation}
where $r$ is the squeezing parameter.

Propagation of each travelling mode is modelled as a thermal-loss channel with transmissivity $\eta$ and environmental variance $\omega$. In covariance-matrix form, each distributed TMSV pair has
the normal form
\begin{equation}
    V_{AB}
    =
    \begin{pmatrix}
        x I & z Z \\
        z Z & y I
    \end{pmatrix},
    \qquad
    I=\mathrm{diag}(1,1),
    \quad
    Z=\mathrm{diag}(1,-1),
    \label{eq:tmsv_cm}
\end{equation}
with
\begin{equation}
\begin{aligned}
    x &= \eta\mu+(1-\eta)\omega,
    \qquad
    y = \mu, \\
    z &= \sqrt{\eta(\mu^2-1)},
    \qquad
    \mu = \cosh(2r).
\end{aligned}
\label{eq:thermal_loss}
\end{equation}
The case $\omega=1$ corresponds to pure optical loss
\cite{Ottaviani2019}.

Because homodyne detection is a Gaussian operation, the conditional output remains Gaussian for any input and can be described entirely by its covariance matrix \cite{Weedbrook2012}. 

In the Ottaviani protocol, the $N$ travelling modes are combined by an $N$-port interferometer comprising $N-1$ cascaded beam splitters with transmissivities
\begin{equation}
    T_k = 1-\frac{1}{k},
    \qquad
    k=2,\ldots,N .
    \label{eq:ott_transmissivity}
\end{equation}
The relay then measures $N-1$ collective position quadratures and one collective momentum quadrature using homodyne detection. The interferometer is chosen such that the conditional state of the retained $B$ modes is invariant under permutation of the users \cite{Ottaviani2019}. For ideal Einstein--Podolsky--Rosen (EPR) resources, the measurement establishes collective momentum and relative-position correlations characteristic of a multipartite CV GHZ/cluster state.

Because both the input states and the homodyne measurement are Gaussian, the conditional output remains Gaussian and is completely
characterised by its covariance matrix. Ottaviani \textit{et al.} derive a closed-form expression for the logarithmic negativity between
any two retained modes,
\begin{equation}
    E_{\mathcal N}^{(N)}
    =
    E_{\mathcal N}^{(2)}
    -
    \frac{1}{2}
    \ln\left(
        1+
        \alpha\frac{N-2}{N}
    \right),
    \label{eq:ott_logneg}
\end{equation}
where
\begin{equation}
\begin{aligned}
    E_{\mathcal N}^{(2)}
    &=
    \ln
    \left[
        \frac{\eta\mu+(1-\eta)\omega}
             {\eta+(1-\eta)\mu\omega}
    \right], \\
    \alpha
    &=
    \frac{\eta(\mu^2-1)}
         {\eta+(1-\eta)\mu\omega}.
\end{aligned}
\label{eq:ott_logneg_parameters}
\end{equation}
This analytic result provides the Gaussian reference against which the non-Gaussian heralding protocols considered below are compared. \\

\section{Multipartite Entanglement Swapping with a DFT--spd Relay}
\label{sec:protocol}

\begin{figure*}[t]
    \centering
    \includegraphics[width=1\linewidth]{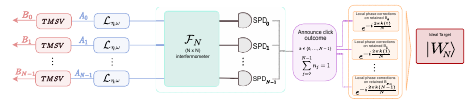}
    \caption  {\small \textbf{Multipartite entanglement-swapping protocol using a balanced DFT relay and single-photon heralding.} Each of the (N) users prepares a two-mode squeezed vacuum (TMSV) state, retaining mode $B_j$ locally while transmitting mode $A_j$ through a thermal-loss channel $\mathcal{L}$ characterised by transmissivity $\eta$ and thermal noise $\omega$. The transmitted modes interfere at a balanced (N)-mode discrete Fourier transform (DFT) relay, followed by single-photon detection. A successful single-photon event at output (k) is classically announced to the users, who apply outcome-dependent local phase corrections to their retained modes. All (N) single-click outcomes are thereby mapped to equivalent symmetric heralded states, with the ideal lossless single-excitation limit given by $|W_N\rangle$.}
    \label{fig:new_protocol}
\end{figure*}

We consider $N$ users sharing independent two-mode squeezed-vacuum (TMSV) states between modes $A_j$ and $B_j$, with $j=0,\ldots,N-1$.
Each user retains mode $B_j$ and transmits mode $A_j$ through a thermal-loss channel to a central relay, $\mathcal{F_N}$. The relay combines the
travelling modes in a balanced discrete Fourier transform (DFT) multiport and performs single-photon-detection (SPD) detection on
each output mode, as illustrated in Fig.~\ref{fig:new_protocol}. 

We first consider the multipartite interferometer of Ottaviani \textit{et al.} \cite{Ottaviani2019}. Replacing the homodyne measurements of that protocol by SPD detection gives one fully symmetric single-click
outcome. In the ideal single-excitation sector, detecting one photon in this symmetric relay mode and vacuum in all remaining outputs
heralds
\begin{equation}
    |W_N\rangle
    =
    \frac{1}{\sqrt{N}}
    \sum_{j=0}^{N-1}
    |1_j\rangle_B ,
    \label{eq:WN}
\end{equation}
where
\begin{equation}
    |1_j\rangle_B
    \equiv
    |0\cdots0\,1_j\,0\cdots0\rangle .
\end{equation}
For $N=3$, this becomes
\begin{equation}
    |W_3\rangle
    =
    \frac{1}{\sqrt{3}}
    \left(
        |100\rangle
        +
        |010\rangle
        +
        |001\rangle
    \right).
    \label{eq:W3}
\end{equation}
The remaining single-click outputs of the cascade relay correspond to contrast modes and produce unequal amplitude distributions across
the retained modes. They are therefore not permutation-equivalent to Eq.~\eqref{eq:WN}. Thus, although SPD detection naturally heralds a symmetric $W_N$ state, only one relay output can be retained if the same multipartite symmetry is required for every accepted event. In these simulations, covariance matrices have been converted to density matrices via The Walrus package by Xanadu \cite{gupt2019walrus}.

Balanced-interferometers approaches have been used for heralded $W$ state distribution, for example the protocol of Roga \textit{et al.} \cite{roga2023efficient} (see also Ref.~\cite{ishihara2026multipartite}). Note that the three-user case of this protocol uses a four mode Hadamard network with an additional vacuum input. Here, we instead employ a native balanced $N$-mode discrete Fourier transform (DFT) relay, avoiding unused vacuum inputs and extra output modes. Its action on the travelling modes is described by

\begin{equation}
    \hat{c}^{\dagger}_k
    =
    \sum_{j=0}^{N-1}
    (F_N)_{kj}\hat{a}^{\dagger}_j,
    \qquad
    (F_N)_{kj}
    =
    \frac{1}{\sqrt{N}}\omega_N^{kj},
    \label{eq:dft}
\end{equation}
where
\begin{equation}
    \omega_N=e^{2\pi i/N},
    \qquad
    j,k=0,\ldots,N-1 .
\end{equation}
Here $\hat{a}^{\dagger}_j$ and $\hat{c}^{\dagger}_k$ denote the creation operators of the DFT input and output modes, respectively.
The DFT is a balanced multiport because
\begin{equation}
    \left|(F_N)_{kj}\right|
    =
    \frac{1}{\sqrt{N}}
\end{equation}
for every $j$ and $k$. Each relay detector therefore receives an equal-magnitude contribution from every transmitted mode, with the
output ports distinguished only by their relative phases. Balanced Fourier multiports with this structure have previously been used in
multipartite linear-optical and photon-counting protocols
\cite{lim2005generalized, su2017multiphoton, guanzon2022ideal}.

The relay accepts measurement patterns containing exactly one photon,
\begin{equation}
    \mathbf{n}^{(k)}
    =
    (0,\ldots,0,1_k,0,\ldots,0),
    \qquad
    k=0,\ldots,N-1,
    \label{eq:single_click_pattern}
\end{equation}
so that detector $k$ records one photon and all other detectors record vacuum. For an ideal TMSV resource,
\begin{equation}
    |\psi_{\mathrm{TMSV}}\rangle_{A_jB_j}
    =
    \sqrt{1-\lambda^2}
    \sum_{n=0}^{\infty}
    \lambda^n
    |n\rangle_{A_j}|n\rangle_{B_j},
    \qquad
    \lambda=\tanh r ,
    \label{eq:tmsv_protocol}
\end{equation}
the transmitted and retained modes have perfect photon-number correlations. Projection of the DFT outputs onto

$\mathbf{n}^{(k)}$ therefore heralds
\begin{equation}
    |\psi_k\rangle_B
    =
    \frac{1}{\sqrt{N}}
    \sum_{j=0}^{N-1}
    \omega_N^{-kj}
    |1_j\rangle_B .
    \label{eq:dft_herald}
\end{equation}
The complex-conjugate phase relative to Eq.~\eqref{eq:dft} results from projecting the transformed relay modes and depends only on the
chosen Fourier convention.

Equation~\eqref{eq:dft_herald} shows the central property of the relay where every single-click result produces the same equal-amplitude
multipartite state. Detector identity changes only the phases of the single-excitation components. For $k=0$, Eq.~\eqref{eq:dft_herald} reduces directly to $|W_N\rangle$.

For $k\neq0$, the phases can be removed using local phase rotations \begin{equation}
    V_k
    =
    \bigotimes_{j=0}^{N-1}
    \exp\left(
        \frac{2\pi i k j}{N}\hat{n}_j
    \right),
    \label{eq:phase_correction}
\end{equation}
such that
\begin{equation}
    V_k|\psi_k\rangle_B
    =
    |W_N\rangle .
    \label{eq:corrected_W}
\end{equation}
The detector outcome $k$ may therefore be broadcast to the users, who either apply the corresponding phase shifts or account for them
in subsequent measurement settings. Since these operations are local unitaries, all $N$ accepted outcomes have identical entanglement.

For the ideal $W_N$ state, tracing out all but two retained modes gives
\begin{equation}
    \rho_{ij}
    =
    \frac{N-2}{N}|00\rangle\langle00|
    +
    \frac{1}{N}
    \left(
        |10\rangle+|01\rangle
    \right)
    \left(
        \langle10|+\langle01|
    \right).
    \label{eq:WN_reduced}
\end{equation}
The corresponding pairwise logarithmic negativity is
\begin{equation}
    E_{\mathcal N}^{W_N}
    =
    \ln\left[
        \frac{
            2+\sqrt{(N-2)^2+4}
        }{N}
    \right].
    \label{eq:WN_logneg}
\end{equation}
Thus the ideal single-excitation herald has a fixed pairwise entanglement for each $N$, decreasing as the excitation is distributed among an increasing number of users.

The protocol is probabilistic. For a general input state after the loss channels, let $\rho_{AB}$ denote the joint state of all
travelling and retained modes and let $U_F$ denote the DFT acting on the travelling subsystem. The unnormalised state conditioned on single-click outcome $k$ is
\begin{equation}
    \widetilde{\rho}^{(k)}_B
    =
    \mathrm{Tr}_A
    \left[
        \left(
            \Pi_k\otimes I_B
        \right)
        U_F \rho_{AB} U_F^{\dagger}
    \right],
    \label{eq:conditional_unnormalised}
\end{equation}
where
\begin{equation}
    \Pi_k
    =
    |\mathbf{n}^{(k)}\rangle
    \langle\mathbf{n}^{(k)}|
    \label{eq:SPD_projector}
\end{equation}
is the SPD projector. The probability and normalised conditional state are therefore
\begin{equation}
    P_k
    =
    \mathrm{Tr}
    \left(
        \widetilde{\rho}^{(k)}_B
    \right),
    \qquad
    \rho_B^{(k)}
    =
    \frac{
        \widetilde{\rho}^{(k)}_B
    }{P_k}.
    \label{eq:conditional_state}
\end{equation}

For identical resources and symmetric channels, the DFT output ports are equivalent and hence all single-click events occur with the same
probability $P_k=P_0$. Since every one of these events is usable after phase correction, the total heralding probability is
\begin{equation}
    P_{\mathrm{succ}}^{\mathrm{DFT}}
    =
    \sum_{k=0}^{N-1}P_k
    =
    N P_0 .
    \label{eq:dft_success_probability}
\end{equation}
In the lossless case, Eq.~\eqref{eq:tmsv_protocol} gives the
closed-form single-port probability
\begin{equation}
    P_0
    =
    (1-\lambda^2)^N\lambda^2,
    \label{eq:single_port_probability}
\end{equation}
and hence
\begin{equation}
    P_{\mathrm{succ}}^{\mathrm{DFT}}
    =
    N(1-\lambda^2)^N\lambda^2.
    \label{eq:ideal_dft_probability}
\end{equation}

This probability enhancement follows from the symmetry of the accepted outcomes rather than an increase in the total number of
single-photon events. With SPD detection applied to the original cascade relay, only the single-click event associated with the
fully symmetric collective mode is retained, whereas the DFT relay allows all $N$ single-click outcomes to be accepted. In the ideal
symmetric case,
\begin{equation}
    \frac{
        P_{\mathrm{succ}}^{\mathrm{DFT+SPD}}
    }{
        P_{\mathrm{succ}}^{\mathrm{Cascade+SPD}}
    }
    =N .
    \label{eq:probability_enhancement}
\end{equation}
For $N=3$, the same mechanism converts one usable herald into three phase-equivalent heralds.

Under channel loss, a successful SPD event may originate from higher-photon-number components of the initial TMSV resources, and the conditional state is therefore generally mixed rather than the pure state of Eq.~\eqref{eq:dft_herald}. We evaluate Eq.~\eqref{eq:conditional_unnormalised} in a truncated multimode Fock basis and calculate the pairwise logarithmic negativity of the normalised retained state. This allows the entanglement and heralding performance of the complete DFT--SPD protocol to be evaluated as a function of squeezing, channel transmissivity, thermal noise, and network size.

\section{Results}
\label{sec:results}

We evaluate the proposed Fourier-relay protocol by comparing three continuous-variable (CV) entanglement-swapping schemes: (i) the Gaussian homodyne protocol of Ottaviani et al.~\cite{Ottaviani2019}, (ii) the original cascade interferometer with single-photon-resolving detection replacing homodyne measurement, and (iii) the proposed discrete Fourier transform (DFT) relay with single-photon-resolving detection.

The comparison separates two effects: the improvement in conditional entanglement obtained through non-Gaussian heralding, and the increase in usable heralding probability enabled by the symmetry of the DFT interferometer.

We consider identical two-mode squeezed vacuum (TMSV) resources with squeezing parameter $r$, transmitted through identical thermal-loss channels characterised by transmissivity $\eta$ and environmental variance $\omega$. Unless otherwise stated, results are presented for $r=0.5$ and $\omega=1$, corresponding to pure-loss channels.

The principal performance metrics are the pairwise logarithmic negativity $E_{\mathcal N}$, the probability
of an accepted heralding event $P_{\mathrm{usable}}$, and the probability-weighted entanglement yield
$Y_E=P_{\mathrm{usable}}E_{\mathcal N}$.

\subsection{Conditional entanglement and loss robustness}
\label{sec:results_entanglement}

We first investigate whether single-photon heralding improves the conditional entanglement generated between remote users relative to the Gaussian homodyne-swapping benchmark.

\begin{figure*}[!t]
    \centering
    \includegraphics[width=1\linewidth]{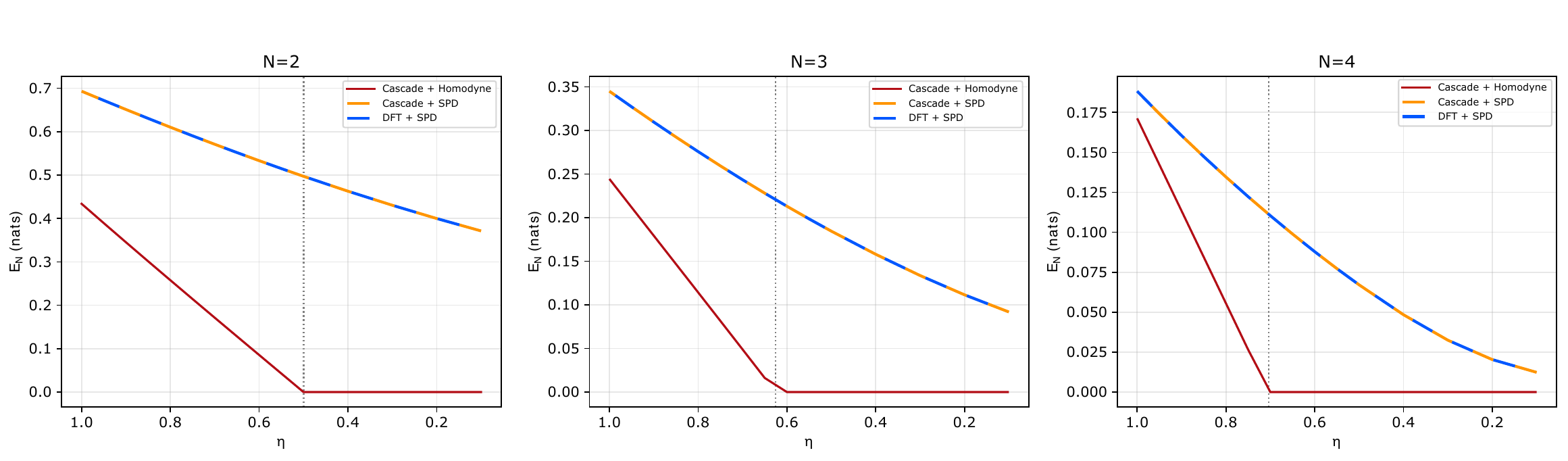}
    \caption{\small
    \textbf{PNR entanglement beyond HOM death ($r$=0.5, $\omega$ = 1)} Pairwise logarithmic 
    channel transmissivity $\eta$ for $N=2,3,4$,with squeezing parameter $r=0.5$ and
    environmental variance $\omega=1$. The Gaussian homodyne benchmark is compared
    with conditional single-photon heralding. The latter preserves non-zero conditional entanglement over the investigated loss range, including regions where the Gaussian-swapped state has zero pairwise logarithmic negativity. 
    }
    \label{fig:entanglement_loss}
\end{figure*}

In the lossless limit, the accepted single-excitation herald produces the symmetric multipartite state $|W_N\rangle$ defined in Eq.~\eqref{eq:WN}. The corresponding pairwise logarithmic negativity
is given by Eq.~\eqref{eq:WN_logneg}. 

For $r=0.5$, the lossless pairwise entanglement obtained from Gaussian homodyne detection and single-photon heralding is

\begin{equation}
\begin{array}{c|ccc}
N & 2 & 3 & 4 \\ \hline
E_{\mathcal N}^{\mathrm{hom}}
  & 0.4338 & 0.2444 & 0.1713 \\
E_{\mathcal N}^{\mathrm{SPD}}
  & 0.6931 & 0.3450 & 0.1882
\end{array}
\label{eq:lossless_comparison}
\end{equation}

The corresponding improvements in conditional pairwise logarithmic negativity are approximately $59.8\%$, $41.2\%$, and $9.9\%$ for $N=2,3,4$, respectively.

The reduction in relative advantage with increasing $N$ reflects the distribution of a single excitation across a larger number of retained modes. Consequently, single-photon heralding does not universally outperform Gaussian homodyne detection. Its advantage depends on the squeezing strength, channel conditions, and number of participating users.

Of particular interest is the behaviour under optical loss. Figure~\ref{fig:entanglement_loss} compares the pairwise logarithmic negativity as the channel transmissivity is reduced from $\eta=1$ to $\eta=0.3$ for $N=2,3,4$.

The Gaussian protocol exhibits a progressive reduction in pairwise entanglement, reaching zero logarithmic negativity at finite transmissivity. By contrast, the single-photon-heralded protocol retains non-zero conditional pairwise entanglement throughout the investigated range.

For example, at $N=3$ and $\eta=0.60$, the Gaussian pairwise logarithmic negativity vanishes, whereas the single-photon-heralded state retains

\begin{equation}
E_{\mathcal N}^{\mathrm{DFT+SPD}}
\simeq 0.213.
\end{equation}

Similarly, for $N=4$ at $\eta=0.70$,

\begin{equation}
E_{\mathcal N}^{\mathrm{hom}}=0,
\qquad
E_{\mathcal N}^{\mathrm{SPD}}\simeq0.110.
\label{eq:n4_loss_example}
\end{equation}

These results demonstrate that single-photon heralding can generate conditional entanglement in lossy regimes where the corresponding Gaussian homodyne protocol no longer generates pairwise entanglement.

Additional thermal-noise sweeps ($\omega>1$) indicate that the relative advantage of single-photon heralding persists beyond pure-loss conditions. Across the investigated parameter grid, PNR detection produced greater conditional pairwise entanglement than Gaussian homodyne detection in approximately $84\%$ of cases at $\omega=1.05$, decreasing to $78\%$ at $\omega=1.20$. The improvement is therefore regime-dependent, with the greatest benefits arising under weak-to-moderate squeezing and lossy conditions. 

\subsection{Symmetric heralding and usable probability}
\label{sec:results_probability}

While single-photon heralding improves conditional entanglement in the lossy regime, its operation is
probabilistic. The probability of obtaining an accepted entangled state is therefore an important measure of the protocol's performance.

The advantage of the DFT relay can be isolated by comparing its accepted detection outcomes with those
obtained by applying the same single-photon detection condition to the original cascade interferometer.

For $N=3$, the cascade interferometer contains a single fully symmetric output mode,

\begin{equation}
M_0 =
\frac{1}{\sqrt{3}}
\left(A_1+A_2+A_3\right),
\label{eq:ott_symmetric_mode_results}
\end{equation}

whose single-photon detection heralds the symmetric $W_3$ state in the lossless case.

The remaining two output modes have unequal-magnitude input coefficients and do not produce equivalent
symmetric heralded states under the same acceptance criterion. Consequently, only the symmetric
detector outcome is retained when the protocol requires the specified permutation-symmetric state.

\begin{figure*}[!t]
    \centering
    \includegraphics[width=1\linewidth]{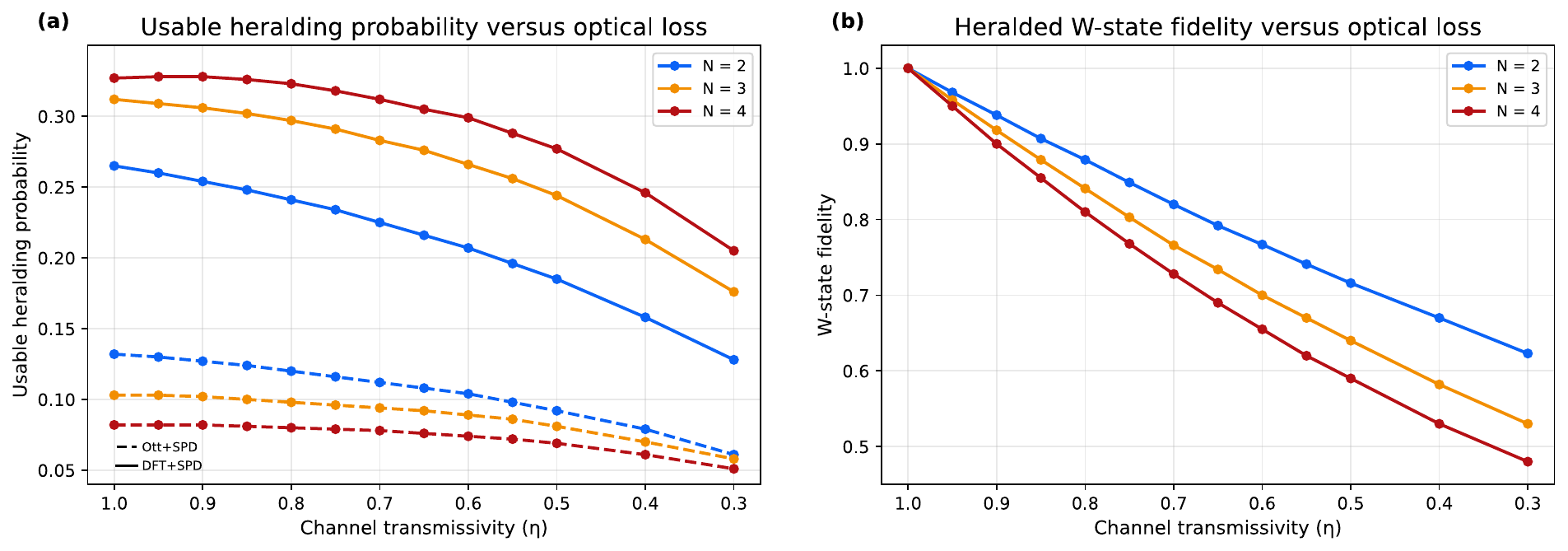}
    \caption{\small
    \textbf{Symmetric heralding advantage of the proposed
    DFT relay for $N=3$.}
    \textbf{(a)} Detector-outcome equivalence: in the original cascade interferometer, only the symmetric collective output mode produces the selected permutation-symmetric single-photon herald. The DFT interferometer produces three equal-magnitude output modes whose accepted single-photon heralds are related by correctable
    local phases. 
    \textbf{(b)} Accepted heralding probability: all three DFT detector outcomes contribute to the usable heralding probability, providing a threefold advantage over the asymmetric relay for the specified single-photon heralding condition.}
    \label{fig:heralding_symmetry}
\end{figure*}



The proposed DFT relay replaces this asymmetric set of output modes with equal-magnitude
superpositions, as described in Sec.~\ref{sec:protocol}. For $N=3$, all three output ports produce
single-excitation states that differ only by known local phases, allowing their accepted
outcomes to be brought to a common form through conditional phase corrections.

We define the usable heralding probability as

\begin{equation}
P_{\mathrm{usable}}
=
\sum_{k\in\mathcal H_{\mathrm{usable}}}P_k,
\label{eq:p_usable_results}
\end{equation}

where $\mathcal H_{\mathrm{usable}}$ is the set
of detector outcomes satisfying the specified
heralding acceptance condition.

At $N=3$, $r=0.5$, $\eta=1$, and $\omega=1$,
the original cascade interferometer with
single-photon detection gives

\begin{equation}
P_{\mathrm{usable}}^{\mathrm{Cascade+SPD}}
\simeq 0.1039,
\end{equation}

whereas the DFT relay gives

\begin{equation}
P_{\mathrm{usable}}^{\mathrm{DFT+SPD}}
\simeq 0.3116.
\label{eq:dft_probability_result}
\end{equation}

Thus,

\begin{equation}
\frac{
P_{\mathrm{usable}}^{\mathrm{DFT+SPD}}
}{
P_{\mathrm{usable}}^{\mathrm{Cascade+SPD}}
}
=3.
\label{eq:probability_ratio}
\end{equation}

Despite this threefold improvement in usable
heralding probability, the conditional pairwise
logarithmic negativity remains

\begin{equation}
E_{\mathcal N}
\simeq 0.3450
\end{equation}

for both protocols at this operating point.

The DFT therefore improves the probability of obtaining the desired symmetric entangled state without reducing its conditional entanglement.

This improvement does not arise from generating three times as many single-photon detection events. Rather, it follows from the equivalence of the DFT detector outcomes, allowing outcomes that do not satisfy the symmetric acceptance condition in the original relay to be retained after known local corrections.

Figure~\ref{fig:heralding_symmetry} (b) further illustrates the dependence of usable heralding probability and $W_N$-state fidelity on channel transmissivity for $N=2,3,4$. The DFT relay increases the usable heralding probability across the investigated network sizes, with an approximately $N$-fold enhancement relative to the cascaded interferometer under the specified acceptance criterion. While the conditional pairwise entanglement remains non-zero under substantial optical loss, the fidelity to the ideal $W_N$ state progressively decreases, with greater degradation observed for larger $N$. This distinction highlights that the persistence of pairwise entanglement does not imply preservation of an ideal multipartite resource state.

\subsection{Probability-weighted entanglement under loss}
\label{sec:results_yield}

The preceding results establish two complementary advantages: single-photon heralding preserves conditional entanglement in lossy regimes where Gaussian homodyne swapping becomes separable, while the symmetric DFT relay improves the acceptance of equivalent heralding outcomes. To quantify their combined operational performance, we define the probability-weighted entanglement yield

\begin{figure}[h]
    \centering
        \includegraphics[width=1\linewidth]{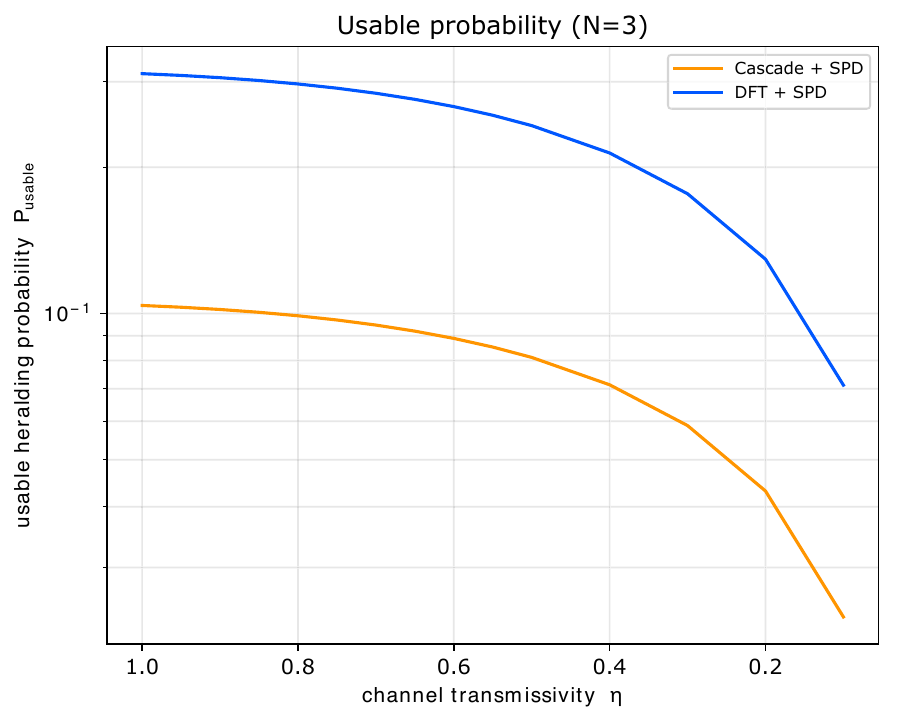}
        \caption{\small
        \textbf{Operational advantage of the symmetric DFT relay under optical loss for $N=3$, $r=0.5$, and $\omega=1$.} Usable single-photon heralding probability for the
        original cascaded interferometer and the DFT relay. }
    \label{fig:yield_loss}
    
\end{figure}
\[
Y_E=P_{\mathrm{usable}}E_{\mathcal N},
\]

where \(P_{\mathrm{usable}}\) is the probability of obtaining an accepted herald and \(E_{\mathcal N}\) is the corresponding conditional pairwise logarithmic negativity.

For \(N=3\), \(r=0.5\), \(\eta=1\), and \(\omega=1\), the cascaded and DFT relay configurations yield

\[
\begin{aligned}
Y_E^{\mathrm{Cascade+SPD}}&\simeq0.0358,\\
Y_E^{\mathrm{DFT+SPD}}&\simeq0.1075.
\end{aligned}
\]

The DFT relay therefore provides a threefold improvement in probability-weighted entanglement yield without changing the conditional entanglement of the accepted state.

Figure~\ref{fig:yield_loss} extends this comparison across channel transmissivity. While the usable heralding probabilities decrease with increasing optical loss, their ratio remains approximately three throughout the investigated range. The conditional pairwise entanglement is likewise preserved between the two relay configurations, with \(E_{\mathcal N}^{\mathrm{DFT}}/E_{\mathcal N}^{\mathrm{Cascade}}\simeq1\). Consequently, the threefold enhancement in probability-weighted entanglement yield persists under loss, demonstrating that the DFT advantage arises from improved utilisation of symmetry-equivalent detection outcomes rather than increased entanglement within individual heralded states.

Although the DFT improves the fraction of usable events, it does not eliminate the probabilistic cost of single-photon heralding, as the absolute success probability decreases with channel transmissivity. The quantity \(Y_E\) therefore provides a comparative measure of operational entanglement yield, rather than a distillable-entanglement rate or quantum communication capacity.

Together, these results demonstrate that non-Gaussian heralding and symmetric interferometric mixing address distinct limitations of multipartite CV entanglement swapping: the former improves conditional entanglement in favourable lossy regimes, while the latter increases the probability of successfully distributing the heralded entangled state.

\section{Discussion and Conclusion}
\label{sec:discussion}

In this work we have we have demonstrated a symmetric Fourier-relay architecture for multipartite continuous-variable (CV) entanglement swapping that combines non-Gaussian single-photon heralding with a discrete Fourier transform (DFT) interferometer. Relative to conventional Gaussian homodyne swapping, the proposed protocol preserves stronger conditional pairwise entanglement in weak-to-moderate squeezing and lossy regimes. The symmetric DFT relay further improves operational efficiency by enabling all single-photon detection outcomes to be accepted up to local phase corrections. For three users, this increases the usable heralding probability and probability-weighted entanglement yield by a factor of three compared with the original asymmetric relay employing the same photon detection, without reducing conditional entanglement.

The improved resilience of conditional entanglement to optical loss comes at the cost of probabilistic heralding, with success probability decreasing as channel transmissivity falls. Performance also depends on squeezing strength and thermal noise, highlighting the regime-dependent nature of the advantage. Nevertheless, the combination of loss-tolerant conditional entanglement and symmetric heralding provides a promising approach to distributing multipartite entanglement across lossy optical networks.

These findings have implications for distributed quantum communication and the generation of multipartite entangled resource states, including larger entanglement networks and cluster-state architectures. By improving the usable entanglement yield, the proposed relay could support more efficient establishment of quantum correlations between remote nodes. Future work should examine scalability to larger networks, realistic detector imperfections, and the preparation and utility of multipartite resource states under practical operating conditions.

\section{Acknowledgements}
J.D. acknowledges support from the Australian Government through an Australian Research Council Discovery Early Career Researcher Award, (Project No. DE250100914). We acknowledge helpful discussions with T. C. Ralph. 

\appendix
%
%

\section{Linear-optical implementation and analytical
characterisation of the DFT relay}
\label{app:dft_relay}

\subsection{Three-mode Fourier interferometer}
\label{app:dft_implementation}

The proposed relay employs a balanced discrete Fourier transform (DFT) interferometer, commonly referred to as a tritter, in which each output mode contains equal-magnitude contributions from all incoming optical modes. The incoming modes \(A_j\), transmitted from the remote users, are transformed into three output measurement modes \(M_k\), each monitored by a single-photon detector. The retained modes \(B_j\) remain at the remote stations and are conditionally entangled following successful single-photon heralding. This contrasts with the cascaded interferometer of Ottaviani \textit{et al.}, whose output modes comprise one fully symmetric collective mode and \(N-1\) orthogonal contrast modes.

For $N=3$, the Fourier transformation is
\begin{equation}
F_3 =
\frac{1}{\sqrt{3}}
\begin{pmatrix}
1 & 1 & 1 \\
1 & \omega_3 & \omega_3^2 \\
1 & \omega_3^2 & \omega_3
\end{pmatrix},
\label{eq:app_f3}
\end{equation}

where $\omega_3=\exp(2\pi i/3)$ satisfies $\omega_3^3=1$ and $1+\omega_3+\omega_3^2=0$.

Denoting the creation operators of the incoming modes \(A_j\) by \(\hat a_j^\dagger\), and those of the output measurement modes \(M_k\) by \(\hat c_k^\dagger\), the Fourier transformation is expressed, using the convention of Sec.~\ref{sec:protocol}, as the output modes satisfy

\begin{equation}
\hat{c}_k^\dagger
=
\sum_{j=0}^{2}(F_3)_{kj}\hat{a}_j^\dagger.
\end{equation}

Explicitly,
\begin{align}
\hat{c}_0^\dagger &=
\frac{
\hat{a}_0^\dagger+
\hat{a}_1^\dagger+
\hat{a}_2^\dagger
}{\sqrt{3}},
\\
\hat{c}_1^\dagger &=
\frac{
\hat{a}_0^\dagger+
\omega_3\hat{a}_1^\dagger+
\omega_3^2\hat{a}_2^\dagger
}{\sqrt{3}},
\\
\hat{c}_2^\dagger &=
\frac{
\hat{a}_0^\dagger+
\omega_3^2\hat{a}_1^\dagger+
\omega_3\hat{a}_2^\dagger
}{\sqrt{3}}.
\end{align}

The equal-magnitude property
\begin{equation}
|(F_3)_{kj}|^2=\frac{1}{3}
\end{equation}
ensures that a photon entering any individual
input mode has equal probability of emerging
at each output port. The transformation therefore distributes each input mode with equal magnitude across all measurement outputs, while the relative Fourier phases distinguish the orthogonal output modes.

The three output modes \(M_0\), \(M_1\), and \(M_2\) are measured independently using single-photon detectors. Successful heralding corresponds to detecting exactly one photon across the three measurement modes, with accepted outcomes \((1,0,0)_M\), \((0,1,0)_M\), and \((0,0,1)_M\). Each outcome projects the retained \(B\) modes onto a conditional state with equal-magnitude single-excitation amplitudes, differing only by relative phases. These phases may be corrected through local operations conditioned on the measurement outcome, allowing all three detection events to contribute to the usable heralding probability.

\begin{figure}[t]
    \centering
    \includegraphics[width=\linewidth]
        {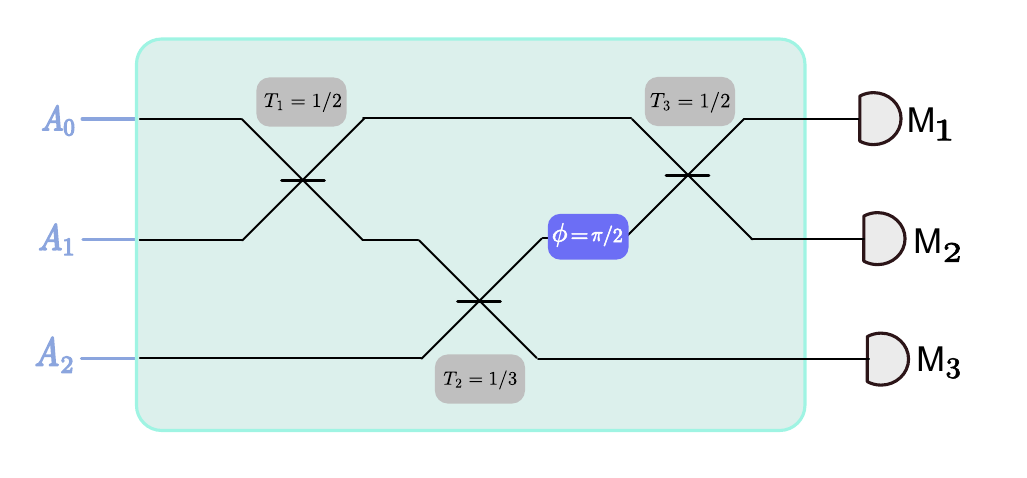}
    \caption{\small \textbf{Linear-optical implementation of the three-mode
discrete Fourier transform (DFT) relay.} The incoming optical modes $A_j$ are mixed using a network of beam splitters and phase shifters, producing the output measurement modes $M_k$. Each output is monitored by a single-photon or photon-number-resolving detector. Successful single-photon heralding corresponds to the outcomes $(1,0,0)_M$, $(0,1,0)_M$, or $(0,0,1)_M$, which generate locally phase-equivalent conditional states on the retained $B$ modes.}
    \label{fig:app_dft_circuit}
\end{figure}

The transformation may be implemented using a passive linear-optical network of beam splitters and phase shifters, as illustrated
schematically in Fig.~\ref{fig:app_dft_circuit}. The beam-splitter transmissivities and relative phases are selected to reproduce the Fourier unitary, up to input/output phase conventions. Individual beam splitters need not all be 50:50 for the complete multiport to be balanced.

\subsection{Single-photon heralded states and phase equivalence}
\label{app:dft_heralded_states}

Consider three independent two-mode squeezed-vacuum resources, with each user retaining mode $B_j$ and transmitting mode $A_j$ to the relay. In the lossless
case, the joint input state is
\begin{equation}
\begin{split}
|\Psi\rangle
= &(1-\lambda^2)^{3/2}
\sum_{n_0,n_1,n_2=0}^{\infty}
\lambda^{n_0+n_1+n_2}
\\
&\times
|n_0n_1n_2\rangle_A
|n_0n_1n_2\rangle_B,
\end{split}
\label{eq:app_joint_tmsv}
\end{equation}
where $\lambda=\tanh r$.

Since passive linear optics preserves total photon
number, projection onto exactly one photon across
the relay outputs selects the single-excitation
sector of the transmitted modes.

Before the relay transformation, this component is
\begin{equation}
\begin{split}
|\Psi_1\rangle
= &(1-\lambda^2)^{3/2}\lambda
\big(
|100\rangle_A|100\rangle_B
\\
&+
|010\rangle_A|010\rangle_B
+
|001\rangle_A|001\rangle_B
\big).
\end{split}
\label{eq:app_single_excitation}
\end{equation}

Following the DFT transformation and projection
onto a single-photon outcome at detector $k$,
the normalised retained state is
\begin{equation}
|W_k\rangle_B
=
\frac{1}{\sqrt{3}}
\sum_{j=0}^{2}
\omega_3^{-kj}|1_j\rangle_B,
\qquad k=0,1,2.
\label{eq:app_conditional_w}
\end{equation}

The three accepted outcomes correspond to
\begin{align}
|W_0\rangle &=
\frac{|100\rangle+|010\rangle+|001\rangle}
{\sqrt{3}},
\\
|W_1\rangle &=
\frac{|100\rangle+\omega_3^2|010\rangle+
\omega_3|001\rangle}{\sqrt{3}},
\\
|W_2\rangle &=
\frac{|100\rangle+\omega_3|010\rangle+
\omega_3^2|001\rangle}{\sqrt{3}}.
\end{align}

Although their relative phases differ, all
three states have equal excitation amplitudes.

Defining the local correction
\begin{equation}
V_k =
\bigotimes_{j=0}^{2}
\exp\left(
\frac{2\pi i kj}{3}\hat{n}_{B_j}
\right),
\end{equation}
gives
\begin{equation}
V_k|W_k\rangle=|W_0\rangle.
\end{equation}

Thus, all three detector outcomes generate
states locally unitarily equivalent to the
symmetric $W_3$ state. Since local unitary
operations preserve entanglement, each
outcome has identical pairwise logarithmic
negativity.

\subsection{Analytical entanglement and heralding probability}
\label{app:dft_analytical_performance}

For the symmetric state $|W_3\rangle$, tracing
out the third retained mode gives
\begin{equation}
\rho_{B_0B_1}
=
\frac{1}{3}|00\rangle\langle00|
+
\frac{2}{3}|\psi^+\rangle\langle\psi^+|,
\end{equation}
where
\begin{equation}
|\psi^+\rangle
=
\frac{|01\rangle+|10\rangle}{\sqrt{2}}.
\end{equation}

The eigenvalues of the partially transposed
density matrix are
\begin{equation}
\operatorname{eig}
\left(\rho_{B_0B_1}^{T_{B_1}}\right)
=
\left\{
\frac{1}{3},
\frac{1}{3},
\frac{1+\sqrt{5}}{6},
\frac{1-\sqrt{5}}{6}
\right\}.
\end{equation}

The corresponding pairwise logarithmic
negativity is therefore
\begin{equation}
E_{\mathcal N}^{W_3}
=
\ln\left(\frac{2+\sqrt{5}}{3}\right)
\simeq0.3450.
\label{eq:app_w_negativity}
\end{equation}

The total probability of obtaining exactly
one photon across the three relay outputs
follows from the norm of the single-excitation
component in Eq.~\eqref{eq:app_single_excitation}:
\begin{equation}
P_{\mathrm{total}}
=
3(1-\lambda^2)^3\lambda^2.
\label{eq:app_dft_probability}
\end{equation}

Fourier symmetry distributes this probability
equally among the three detector outcomes,
\begin{equation}
P_{100}=P_{010}=P_{001}
=
(1-\lambda^2)^3\lambda^2.
\end{equation}

For $r=0.5$, the corresponding values are
\begin{align}
P_{\mathrm{branch}}&\simeq0.103876,\\
P_{\mathrm{total}}&\simeq0.311627.
\end{align}

Accepting all three outcomes therefore triples
the usable heralding probability relative
to retaining a single symmetric branch.

Since the conditional entanglement is identical
for all three phase-corrected outcomes, the
probability-weighted pairwise entanglement yield
also increases by a factor of three.

\subsection{Comparison with the cascaded relay}
\label{app:cascade_comparison}

The origin of the probability advantage can
be understood directly from the output-mode
transformations.

The original cascaded relay contains one
fully symmetric collective mode,
\begin{equation}
\hat{c}_{\mathrm{sym}}^\dagger
=
\frac{1}{\sqrt{3}}
\left(
\hat{a}_0^\dagger+
\hat{a}_1^\dagger+
\hat{a}_2^\dagger
\right).
\label{eq:app_cascade_symmetric}
\end{equation}

This is identical, up to phase convention,
to the $k=0$ output mode of the
three-mode DFT.

For a measurement requiring one photon
in the symmetric mode and vacuum in all
orthogonal relay modes, the corresponding
input-space projection is identical for
the two interferometers.

Consequently, for the same input state
and channel conditions, the selected
conditional retained state and its
heralding probability are the same for
the symmetric cascade branch and the
$k=0$ DFT branch.

The remaining cascade outputs correspond
to orthogonal contrast modes with
unequal-magnitude input coefficients.
Their single-photon projections generally
produce states that cannot be converted
into the specified symmetric $W_3$
state by local phase rotations.

By contrast, all three DFT output modes
have equal-magnitude coefficients and
generate locally phase-equivalent states
in the ideal lossless case.

Under the specified symmetric-state
acceptance criterion, the DFT therefore
provides three usable single-photon
outcomes compared with one for the
original cascade relay:
\begin{equation}
\frac{
P_{\mathrm{usable}}^{\mathrm{DFT+PNR}}
}{
P_{\mathrm{usable}}^{\mathrm{Cascade+PNR}}
}
=3.
\label{eq:app_probability_ratio}
\end{equation}

The improvement arises from accepting
additional equivalent measurement outcomes
rather than increasing the conditional
entanglement of an individual heralded
state.

\newpage

\section{Numerical methodology and validation}
\label{app:numerical}

\subsection{Gaussian state preparation and channel modelling}
\label{app:gaussian_preparation}

The numerical calculations begin with $N$ independent
two-mode squeezed-vacuum (TMSV) states, each comprising
a travelling mode $A_j$ and a locally retained mode
$B_j$. The squeezing parameter $r$ determines the
initial correlations between the two modes.

In the quadrature convention where the vacuum variance
is unity, the covariance matrix of each TMSV pair is
\begin{equation}
V_{\mathrm{TMSV}} =
\begin{pmatrix}
\mu I & cZ \\
cZ & \mu I
\end{pmatrix},
\end{equation}
where
\begin{equation}
\mu=\cosh(2r), \qquad
c=\sinh(2r),
\end{equation}
and $I=\mathrm{diag}(1,1)$ and
$Z=\mathrm{diag}(1,-1)$.

The travelling modes propagate through independent,
identical thermal-loss channels characterised by
transmissivity $\eta$ and environmental variance
$\omega$. The resulting covariance matrix for each
distributed pair takes the normal form
\begin{equation}
V_{A_jB_j} =
\begin{pmatrix}
xI & zZ \\
zZ & yI
\end{pmatrix},
\end{equation}
with
\begin{equation}
\begin{aligned}
x &= \eta\mu+(1-\eta)\omega,\\
y &= \mu,\\
z &= \sqrt{\eta}\,c.
\end{aligned}
\end{equation}

The case $\omega=1$ corresponds to pure optical loss.
For $\omega>1$, the environment introduces additional
thermal noise.

The combined covariance matrix is constructed as
the direct sum of the $N$ distributed resource
pairs, with the mode ordering chosen consistently
with the subsequent interferometric transformation.

The relay transformation acts only on the travelling
$A$ modes. For a passive interferometer represented
by a symplectic matrix $S_R$, the covariance matrix
transforms according to
\begin{equation}
V_R=S_RV_{\mathrm{in}}S_R^T.
\end{equation}

The Gaussian homodyne benchmark is evaluated using
the analytical pairwise logarithmic-negativity
expressions of Ottaviani \textit{et al.}
The photon-number-heralded protocols instead require
a density-matrix representation of the state after
the relay transformation.

\end{document}